\documentclass{aa}  
\usepackage{graphicx}
\usepackage{txfonts}
\def\lya{Ly$\alpha$ }
\def\lbgn{$LBG_N$ }
\def\lbgl{$LBG_L$ }
\begin{document}
   \title{Physical properties of z$\sim$ 4 LBGs: differences between galaxies with and without Ly$\alpha$ emission}


   \author{   L.Pentericci\inst{1}
             \and 
          A. Grazian \inst{1}
          \and
          A. Fontana \inst{1}
          \and
          S. Salimbeni \inst{1}
          \and
         P. Santini \inst{1}
          \and
          C. De Santis \inst{1}
          \and
          S. Gallozzi \inst{1}
          \and
          E. Giallongo \inst{1}
          }

   \offprints{L.Pentericci}

   \institute{INAF - Osservatorio Astronomico di Roma, Via Frascati 33,
I--00040, Monteporzio, Italy
             }

   \date{}

 
  \abstract
   {}
   {We have analysed  the physical properties of z $\sim$ 4 Lyman Break Galaxies  observed in  the GOODS-S survey, in order to  investigate the possible differences between galaxies where the Ly$\alpha$ is present in emission, and those where the line is absent or in absorption.}
   { The objects have been selected from their optical color and then spectroscopically confirmed by  Vanzella et al. (2005). From the public spectra 
we assessed the nature of the \lya emission and  
 divided the sample into galaxies with \lya in emission 
and objects without \lya line (i.e. either absent or in absorption).
We have then  used the complete photometry, from U band to mid infrared from the GOODS-MUSIC database, to study  the observational properties of the galaxies, such as UV spectral slopes and optical to mid-infrared colors, and the possible differences between the two samples.

Finally through standard spectral fitting tecniques we have determined 
 the physical properties of the galaxies, such as total stellar mass, stellar ages and so on, and again we have studied the possible differences between the two samples.}
   {Our results indicate  that LBG with \lya\ in  emission  are on average a much younger and 
less massive population than the  LBGs without \lya emission. Both populations are forming stars very actively and are relatively dust free, although those with line emission 
 seem to be even less dusty on average. We briefly discuss these results in the context of recent  models for the evolution of Lyman break galaxies and \lya emitters.}
   {}


\maketitle


\section{Introduction}

In the past years large samples of   galaxies
 have been found up to the highest redshifts (Iye et al. 2006, Kashikawa et al. 2006, Bouwens et al. 2004) using techniques that
 rely on various color-selection criteria. 
 Among the various methods, one of the most efficient is the Lyman break dropout technique (Steidel \& Hamilton 1993) which  is sensitive to the presence of the 912 \AA\  break and
is effective in finding star
forming galaxies.  This method requires a  blue spectrum, implying low to moderate dust absorption. It  was first 
 designed to select $\sim 3$  galaxies (Steidel et al. 1996, Madau et al. 1996) with a typical magnitude limit R$>$25.5, 
and then extended  to higher redshift (Steidel et al. 1999; Ouchi et al. 2004; Giavalisco et al. 2004; Dickinson et al. 2004). 
\\
 An alternative technique to find such distant objects is to search 
for \lya\ emission,  through very deep, narrow band imaging at selected redshift windows,  as first shown  by Cowie \& Hu (1998).
\lya emitters (LAEs) are generally selected to have high restframe \lya\ equivalent width, typically $EW \gtrsim 20 \AA$, with no constraint on the continuum. 
 Therefore, this method tends to select much fainter galaxies, compared to the general LBGs population. Many \lya emitters have now been found 
(e.g. Iye et al. 2006, Ouchi et al. 2004, Fujita et al.2003), and  several 
distant large scale structures or protoclusters  have been discovered 
(e.g. Ouchi et al. 2005, Venemans et al. 2007).
\\
Each of the two methods suffers from different selection bias:  the two resulting galaxies population are partially overlapping, 
and it is not clear what is the  relation between them.
\\
Various scenarios have been proposed to explain the properties of Ly$\alpha $ emitters. Based on rest-frame optical photometry of LBGs, Shapley et al. (2001) concluded that LBGs with Ly$\alpha $ in emission are "old'' (ages larger than  few $\times 10^8$ yr), while "young'' (ages less than $\sim$100 Myrs) LBGs have Ly$\alpha $ in absorption. This could be explained if the  young galaxies contain 
dust which absorbs the  Ly$\alpha $ photons, while the older galaxies are more quiescent with less dust and superwinds which allow the Ly$\alpha $ photons to escape.
\\
In alternative, other groups have suggested that strong Ly$\alpha $ emitters are instead 
 young star forming galaxies, as  derived from the blue colors and high equivalent widths of the Ly$\alpha $ emitters (e.g. Le Fevre et al. 1996, Malhotra  $\&$ Rhoads 2002; Rhoads $\&$ Malhotra 2001; Tapken et al. 2004; Keel et al. 2002).
\\
Finally some authors have suggested that galaxies could have more than one 
\lya bright emission phase (e.g. Thommes \& Meisenheimer 2005). An initial - primaeval - phase in which dust is virtually non-existent, and a later secondary phase in which strong galactic winds as observed in some Lyman break galaxies facilitate the escape of Ly-${\alpha } $ photons after dust has already been formed.
\\ 
Clearly, it  would be important to understand the real relation between galaxies with \lya emission and the general LBG population,
 so that properties of the overall high redshift galaxy population, such as the total stellar mass density, can be better constrained.
\\
To asses this issue we have analised a sample of LBGs 
selected as B-dropouts from the GOODS-South  sample, 
and with VLT spectroscopic confirmation (Vanzella et a. 2005, Vanzella et al. 2006).
Given the good quality of the spectra it was possible to asses 
whether the galaxies  have 
\lya in emission 
or the line is  absent and/or in absorption.
We have then analised how the observed properties and the derived 
physical properties such as total stellar mass and age depend 
on the nature of the Ly$\alpha$ emission.
\\
All magnitudes are in the AB system (except where otherwise stated)
and we adopt the
$\Lambda$-CDM concordance cosmological model ($H_0=70$, $\Omega_M=0.3$ and
$\Omega_{\Lambda}=0.7$).

\section {Sample and observational properties}
From the GOODS-S public data survey we have selected all galaxies with spectroscopic redshift in the range $3.4< z< 4.8$. These galaxies were initially selected as B-band (or V band for those at $z \ge  4.5$) dropouts by Giavalisco et al. (2004) (see this paper for the color selection criteria adopted) and with a z-band constrain $z<26$.
Spectroscopic observations were carried out  with FORS2 in the frame of the GOODS project (Vanzella et al. 2005, Vanzella et al. 2006). 
The lower redshift cut is simply given by the actual 
availability of FORS2 spectra, given that the \lya\ line falls at $\sim$ 5300 \AA, which is at the limit of efficiency for the observation-setup used by Vanzella et al. (2006) (FORS2 plus the 300I grism)\footnote{At lower redshift spectroscopic observations have been carried out with VIMOS but data are still not available to the public.}; the highest redshift cut was chosen 
since at redshift higher than $\sim$ 4.8 , 
the spectroscopic confirmations  are almost exclusively 
based on the presence of the Ly$\alpha$ line in emission, 
while the objects with possible Ly$\alpha$ in absorpion (or absent) 
become progressively more difficul to identify. In fact the most distant 
object identified with Ly$\alpha$ in absorption is at z$=$4.788.
\\ 
From the public GOODS data we have retrieved the spectra, 
of quality A and B. The spectra with quality flag C were only included 
if our independently measured photometric redshift (from 14 bands photometry, 
Grazian et al. 2006)  was in agreement with the spectroscopic redshift, within the errors.
In the GOODS survey spectra are classified as emitters or absorbers, depending on the nature of the \lya line. 
After removing AGNs, in the above redshift range there are 47 galaxies, of which  19 have the line in emission, 
and 28 have Ly$\alpha$ in absorption or absent.
The implied ratio of line emitters/total is quite similar  
to the proportion of Ly$\alpha$ emitters  
that  was found at redshift $z \sim$3 in a large sample of LBGs by Steidel et al. (2000).
We will call the first sample $LBG_L$ (LBG with Line emission) and the second sample  $LBG_N$ (LBGs with No line).
\\
In Figure 1 (upper panel) we show the redshift 
distributions of the two samples, which are similar.
From our GOODS-MUSIC database  we have then gathered the complete  
14 bands multicolor information, 
extending from the U band to the Spitzer $8 \mu$m band 
(for details of the catalog see Grazian et al. 2006).
We have then studied the observational properties: in particular we have derived the restframe luminosity at 1400 \AA, the UV continuum slope $\beta$ and the z-[3.6] color which enconpasses the 4000 \AA\ break. The average observational properties of the two samples, as well as the physical properties that will be derived in the next section,  are reported in Table 1, together with the uncertainties, derived using  the standard deviation of the mean. In the last column of the Table we report, for each parameter, the probability  value P  given by the Kolmogorov-Smirnov test (KS-test) which tries to determine if the two datasets differ significantly. The KS-test has the advantage of making no assumption about the distribution of data.
\\ 
The UV continuum slope was derived from the $i-z$ color,
  which spans a restframe wavelength range from $\sim 1500 \AA$ 
to $\sim 1800 \AA$ (at the mean redshift of the sample $z \sim 4$, and considering the central wavelengths of the ACS filters). 
Although the restframe wavelength baseline is not large  (and perhaps a $i-J$ color would be preferred),  we have used the $i-z$ color for two reasons: for a very few  galaxies we do not  have J band information available 
(see Grazian et al. 2006 for details); second, the UV slope derived in such way 
is more easily comparable to other UV slope values in the literature. 
 Like other authors, we assume  a standard power law spectrum with slope 
$\beta$ ($f_\lambda \propto \lambda^{\beta}$, so that a spectrum that is flat in $f_\nu$ has $\beta =-2$ ). The 1216 \AA\  break only starts to enter the i filter at z$>$4.7 so
 we neglect the impact of neutral hydrogen absorption for all galaxies. 
The measured slopes are plotted in Figure 1 (lower panel) and range  between 
-1 and  -2.5 (with few exeptions).  The uncertainties  on the determined 
 values 
 range from 0.2 for the brightest objects to 0.8 for the fainter ones: in the figure we have indicated a median error on the individual values of 0.5 in the upper right corner.
The average value is $\langle \beta \rangle=-1.8\pm 0.13$  for the overall sample, 
and $\langle 
\beta \rangle=-2.0 \pm 0.11$  and $-1.7\pm 0.13$ considering the $LBG_L$ and $LBG_N$ separately.
The UV slopes we find are consistent with the prediction of models for unobscured continuosly star forming galaxies (e.g. Leitherer et al. 1999). They are also consistent with what found by other authors at a similar  redshift: for example  Overzier et al. (2006), for redshift z$=$4.1 LBGs and LAEs, find  $\beta =-1.95$ , Venemans et al. (2005) find $\beta =-1.65$ for \lya emitters at z$=$3.1. The values are also  similar to the average 
$\beta = -1.8 \pm 0.2$ of V606 dropouts in GOODS found by Bouwens et al. (2006).
\\
On the other hand  our values are a bit steeper than what found by  Papovich et al. (2001) for LBGs at z$=$3 ($\langle \beta \rangle=-1.4$).
Finally Shapley et al. (2003), found less steep values and claim a
 considerable  difference between the $LBG_L$ and $LBG_N$ at z$\sim$3, with the former having a steeper slope, with a positive dependence of slope  on the 
\lya\ equivalent width. 
\\
 As can be seen in Figure 1, the values of the UV slopes  for \lbgl\ and \lbgn\  are largely overlapping and a K-S does not reject the hypothesis that the two samples could be
drawn from the same underlaying distribution (the probability P is at $\sim 1 \sigma$ see Table 1).
However  the median values are indeed somewhat different, and the trend is the same as what found by Shapley et al. (2003)  given that the $LBG_L$  are   
bluer on average than the  $LBG_N$.  
The difference between our two samples  ($\Delta \beta \sim  0.3$) 
 is actually similar  to that found by  Shapley et al. (2003), who have a $\Delta \beta = 0.36$ between their most extreme groups i.e. the 
strong emitters and the strong absorbers (see Table 3 of the mentioned paper), but the significance of their  $\Delta \beta$ is higher, given that they have a much larger sample with 800 objects in total, and therefore the uncertainties on the average $\beta$ values are much lower than in our case.  
In conclusion, although we cannot claim such a strong dependence of $\beta$ on the 
\lya\ properties, we do indeed find a mild indication that \lbgl\ are bluer than \lbgn.
To support this result, a similar difference is also found for 
LBG galaxies by Vanzella et al. (2006),  
who performed spectroscopy of B dropouts from the GOODS sample, and then stacking the spectra of galaxies with \lya\ in emission and in absorption separately 
found that the former have a systematically bluer 
UV continuum than the latter. 
\\
We find no evidence for a slope-magnitude relation  as found by other authors 
(e.g. Overzier et al. 2006 ) but  the errors on the slope are quite large and the wavelength baseline for the slopes is rather small.
\\
In Figure 2
we show the $z-[3.6]$ color plotted versus the total $[3.6]$ magnitude:
this color encompass the 4000\AA /Balmer  breaks, and is sensitive both to dust and stellar ages. The $[3.6]$ magnitude 
corresponds to a rest-frame optical magnitude. 
The strong correlation between the optical flux and color is 
due to the magnitude cut at  $z=26$ that was used for the sample (see dotted line).
It is clear from the figure that the \lbgn\ are on average both brighter 
(by $\sim 0.7$ magnitudes) at  restframe optical wavelengths  and redder (by more than half 
a magnitude)  than the \lbgl.
 Given that all objects are  relatively dust free 
(as inferred from the UV spectral slopes and 
consistently with the Lyman Break selection criteria), 
these differences indicate  that the \lbgn\ have in general a more evolved 
stellar population and they are more massive.
We will derive the physical properties in the next section.
\begin{figure}
\includegraphics[width=9cm]{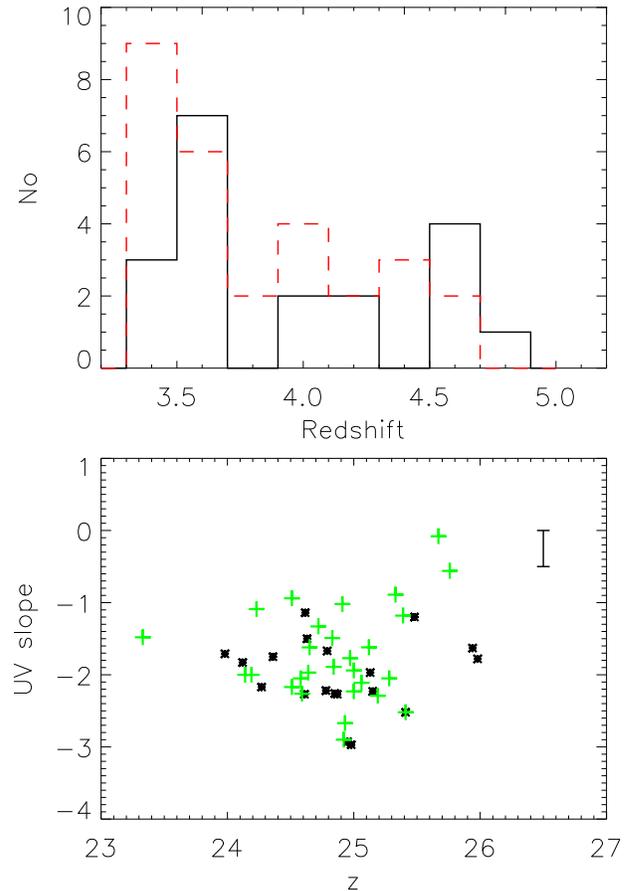}
\caption{Upper panel: the redshift distribution of the two samples. Full black line are the \lbgl, light dashed line are the \lbgn.
Lower panel: the UV slope calculated from the i-z colors, squares are the \lbgl\ and plus signs are the \lbgn.
}
\label{fig1}
\end{figure}
\section{Physical properties}
\subsection{Method}
Using the multiwavelength GOODS-MUSIC data and the spectroscopic redshifts, 
we have then  determined the
 physical properties of these galaxies, through a spectral fitting technique.
The spectral fitting technique adopted here is the same
that has been developed in previous papers
(Fontana et al. 2003, Fontana et al. 2006), and similar
to those adopted
by other groups in the literature (e.g. Dickinson et al. 2003, Drory et al. 2004).
Briefly, it is based on the comparison between the observed multicolor
distribution of each object and a set of templates, computed with standard
spectral synthesis models (Bruzual \& Charlot 2003 in our case),
and chosen to broadly encompass the variety of star--formation histories,
metallicities and extinction of real galaxies.  To compare with
previous works, we have used the Salpeter IMF, ranging over a set of
metallicities (from $Z=0.02 Z_\odot$ to $Z=2.5 Z_\odot$) and dust
extinction ($0<E(B-V)<1.1$, with a Calzetti or a Small Magellanic Cloud
extinction curve). Details
are given in Table 1 of Fontana et al. (2004). 
For each model of this grid, we have
computed the expected magnitudes in our filter set, and found the
best--fitting template with a standard $\chi^2$ minimization. The
stellar mass and other best--fit parameters of the galaxy,
like SFR estimated from the UV luminosity  and corrected for dust obscuration (with a typical correction factor of $A_V \sim 0.4$), age, $\tau$ (the star formation e-folding timescale),
metallicity and dust extinction, are fitted simultaneously
to the actual SED of the observed galaxy.
The derivation of these parameters is explained in detail 
in the above paper and in Fontana et al. (2006), where 
also the uncertainties are discussed.
In particular we note here that  the
stellar mass generally turns out to be the least sensitive to
variations in input model assumptions, and the extension of the
SEDs to the IRAC mid-IR data, 
tends to reduce considerably the formal uncertainties
on the derived stellar masses. 
On the other hand, the physical parameter with highest associated uncertainty is the metallicity, given that the models are strongly degenerate when 
 fitting broad band SEDs;
a further limitation is that  only 4 values were allowed in the modelling, i.e. 0.2,0.02, 1 and 2.5 (respectively subsolar, solar and supersolar metallicity), but in our samples none of the galaxies turned out to have a metallicity of 2.5.
\subsection{Results}
In Figure 3 we plot the distribution of the four basic physical parameters 
for the two samples, namely the total stellar mass, the star formation rate, the derived stellar age and the E(B-V).
In each figure we also indicate with an arrow the average values calculated
for each sample.

Both the total stellar masses  and the median ages are  considerably 
lower for the \lbgl\ compared to the \lbgn.
The average mass is $(5.0 \pm 1)\times 10^9 M_{\odot}$ for the \lbgl\ and  
$(2.3 \pm 0.8) \times 10^{10}  M_{\odot}$
for the others, i.e. a factor of almost five  higher.  
The K-S test gives a very low value, implying that the two 
populations are different from each other  
with $>99.8 \%$  probability. We can therefore conclude that the 
\lbgl\ are a factor of almost 5 less massive than the \lbgn.
\\
The median ages are also quite different, with an average  of $200\pm 50$ Myr 
for the \lbgl,  an age  distribution that  is very peaked 
towards small ages and is  basically confined to values  below 300 Myr.
\begin{figure}
\includegraphics[width=8.6cm]{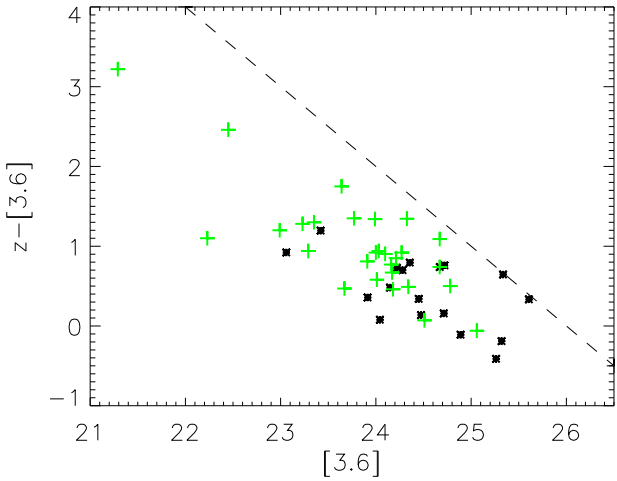}
\caption{The z-[3.6] color plotted vs. observed flux in [3.6].
Squares are the \lbgl\ and plus signs are the \lbgn.}
\label{z1}
\end{figure}
The \lbgn\ on the other hand have an average age of $410\pm 70$ Myr 
(a factor $ \ge $2  higher)
and there are several galaxies with ages exceeding 1 Gyr, which is a considerable fraction of the cosmic time at redshift $\sim 4$.
Again performing a K-S test,  the two populations are 
different with a  probability $>98 \%$ .
We can conclude that the \lbgl\ are significantly younger galaxies compared to the \lbgn\.
\\
The star formation rates do not differ much for the two samples,   
with median values that are almost equal, $\langle SFR \rangle  =76\pm 34 M_{\odot} yr^{-1}$ and $74\pm 22 M_{\odot} yr^{-1}$ respectively and very similar distributions.
The exctinction values are slightly higher for the \lbgn\,  with $\langle E(B-V) \rangle  =0.10\pm 0.02$, while the \lbgl\ have $\langle E(B-V) \rangle =  0.07\pm 0.01 $.
In both cases the K-S test does not reject 
the hypothesis that the 2 groups are drawn from the same population of galaxies, although the average E(B-V) is higher for the \lbgn. 
Therefore both \lbgl\ and \lbgn\ are galaxies with relatively little dust content,  and significant star formation rates.  
In particular the average E(B-V) is quite lower than the average values found by Shapley et al (2003) at redshift $\sim$3 and Ouchi et al. (2004) at redshift $\sim$4  but we find the same tendency for \lbgn\ to have somewhat higher dust content compared to \lbgl\, which is also supported by the difference in the UV  spectral slopes discussed in the previous section.
\\
The final two physical parameters that come as output 
from the modelling  are the metallicity, and $\tau$ i.e.
the star formation e-folding timescale.
Each of them suffers from larger uncertainties as compared to e.g. the 
 total stellar mass, as already discussed before.
The mean values and the total distribution for 
these two parameters are very similar both for $LBG_N$ and $LBG_L$, 
with the K-S test  indicating that they are drawn from the same population.
A more interesting and better constrained 
quantity to determine is the $age/\tau$ parameter 
that is an indication of the evolutionary state of the galaxies. 
It can be shown that passively evolving galaxies can be selected 
according to the physical criterion $age/\tau\ge 4$ (Grazian et al. 2006).
All galaxies in our sample, both $LBG_N$ and $LBG_L$ have  $age/\tau \le 4$
as expected: however the difference in age discussed before 
is reflected also on the difference in   $age/\tau$,  which is relatively 
larger for  $LBG_N$ with a mean value 
of $ \langle age/\tau\rangle_{LBG_N}=0.99$ while for  $LBG_L$ the mean value is 0.33.
Furthermore all but one  $LBG_L$ have  $age/\tau \le 1$
while amongst  $LBG_N$ there are 7 objects with values between 1 and 4.

From this data we conclude that the \lbgl\ seem to be a population 
of younger and less massive galaxies. Given that all galaxies have relatively low dust content, there are some indication that \lbgl\  might be  relatively less dusty than \lbgn\ in general, based on the results presented here and by other authors (Vanzella et al. 2006, Shapley et al. 2003 and Overzier et al. 2006).
The star formation rates are high in all cases 
and are independent on the nature of the \lya emission. 
\begin{figure*}
\center{
\includegraphics[width=15cm]{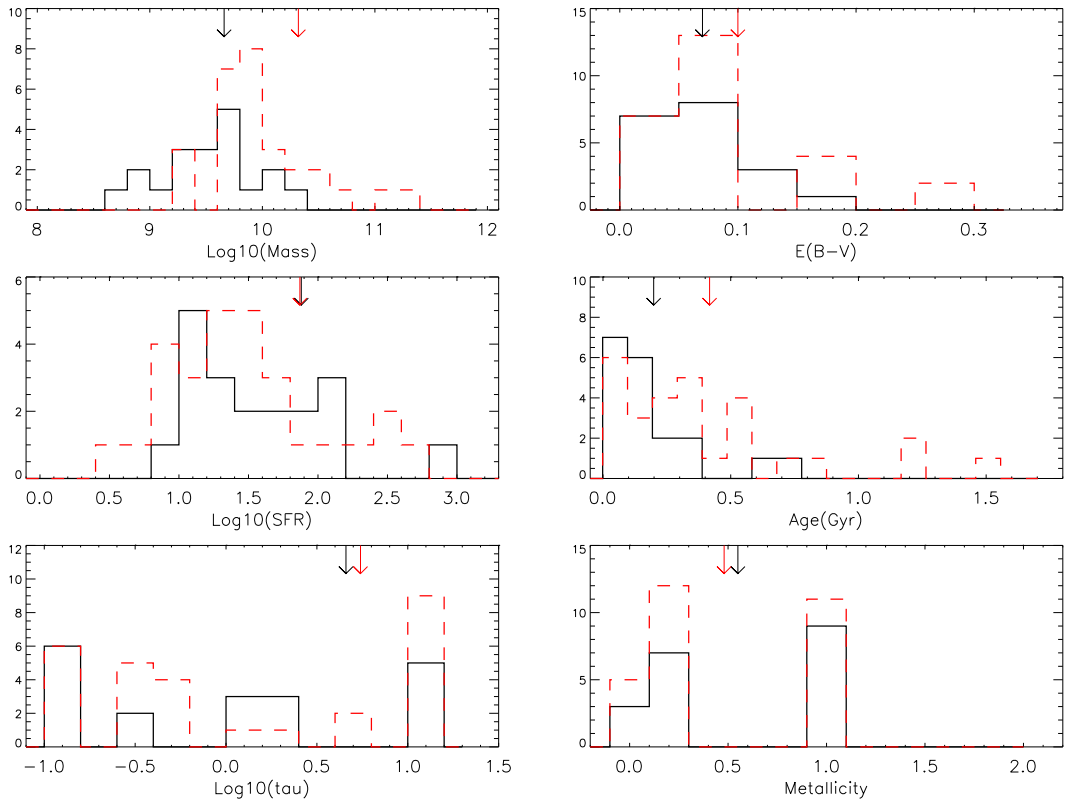}
}
\caption{
The physical properties of B  and V-dropouts. Top left: distribution of total stellar mass; top right the exctinction parameter E(B-V) derived from the spectral fitting; middle left the total star formation rate derived from the spectral fitting; middle right: stellar ages; bottom left: $\tau$ the star formation e-folding time-scale; bottom right: the metallicity. Solid black and dashed light histograms denote \lbgl\ and \lbgn\, respectively. The blck and light arrows indicate the mean values of each sample.}
\label{z2}
\end{figure*}
\section{Discussion and Conclusions}
The results obtained here seem to point towards a scenario where 
\lbgl\ are young and relatively 
small galaxies compared to the general LBGs population, are basically unobscured and are forming stars at very high rates.
$LBG_N$ on the other hand, are more massive, with masses as high as 10$^{11} M_\odot$, and span a larger range of ages from few tens of Myrs to more than 1 Gyr. They are forming stars at comparable rates and are also relatively unobscured galaxies, although they appear slightly redder than \lbgl.
This is therefore partially at variance with the only comparable large  study, carried out at redshift $\sim$3 by Shapley et al. (2003), who found a tendency for \lbgl\ to have in general higher ages, lower values of E(B-V) and higher star formation rates. A very similar work by Iwata et al. (2005) on z$\sim 3$ LBGs from the Hubble Deep Field-South,  also examined the relationship between spectroscopic features and parameters of the best-fit stellar population models in a sample of spectroscopically confirmed galaxies. They did not find any clear correlation between the Ly$\alpha $ emission, Ly$\alpha $ equivalent widths and the results of SED fitting. However the absence of any clear trend may be attributed to the very  small number of sample galaxies in their study (13 in total).
\\
 Some other observational studies seem to support our results: 
for example Overzier et al. (2006) find similar trends for a small 
sample of LBGs and LAEe in a z$=$4.1 protocluster, with indications that  
LAEs are young, dust free objects with blue UV optical colors and 
possibly  less massive that  UV selected LBGs with comparable star formation rates. 
\\
At higher redshift, Lai  et al. (2006) constrained the ages of \lya emitters  
at redshift 5 and found typical ages 50-100 Myr, i.e. comparable to our \lbgl.
Finally at  redshift 3, Gawiser et al. (2006) analised a sample of narrow band selected \lya emitters at z$\sim$ 3 and concluded that they appear to have much less dust and lower stellar masses, compared to the generic LBG population at the same redshift. They were not able to constrain ages, which had a large range of allowed values.
However given that they started from a narrow band sample, their objects are in general much fainter  than the general LBGs population at the same redshift 
and clearly they are not selected to have the same broadband optical colors as the LBGs at the same redshift.
In our case the optical colors and absolute restframe UV luminosity 
of the \lbgn\  and \lbgl are equal.

Our results give support to models that 
consider the \lya\ emitters as precursors of spheroid galaxies,  
observed during the initial starburst phase, which is confined to 
a short period after its onset, due to the rapid formation of dust.
Interestingly one such model (Thommes \& Meisenheimer 2005) attempts to constrain the duration of this initial phase from observational data, finding 
$T_{Ly\alpha} \le 350$ Myr.
A very similar results was recently 
found by Mori \& Umemura (2006) from simulation with a 
N body/hydrodynamics code: they  follow the early stages of galaxy 
formation and find a first phase characterised by strong \lya emission, which after 3$\times 10^8$ years  quickly declines to fluxes below the observable 
level. The ages found in these models, are very  similar to the range of 
allowed ages for our \lbgl\, as shown in Figure 3; 
on the other hand  many of the \lbgn\ have ages exceeding 300 Myr. 
Clearly there are also very few   
\lbgl\ with larger ages and most of all, there are several 
\lbgn\ which are very young, with ages of a few tens of Myrs. 
The first few outliers 
could be easily accounted for 
 given that the modelled physical parameter are obviously 
subject to uncertainties. However the very young \lbgn\
i.e. primeval objects without \lya emission are hard to fit within the simple scenario proposed above.
Probably more complex models in which the time scales for dust 
formation  in primeval galaxies 
is variable and depends on other factors have to be invoked.
In this context, several scenarios have been proposed, for example 
where  the  differential  attenuation 
of the \lya\ and continuum photons depends on the  clumpyness of the  medium 
(e.g. Hansen \& Oh 2006), or where the timescales for the \lya\ emission 
  varies for galaxies residing in halos of different sizes, in the sense that in the most massive halos the \lya luminosity declines abruptly after a much  
shorter time ($10^7$ ys) than in less massive halos (Mao et al. 2007). 
However a full discussion of the models is beyond the scope of this paper.

Finally it also remains unclear whether the \lbgn\ are actually dustier than the \lbgl, which would be a natural consequence of this model: 
we do indeed find that \lbgl\ 
are slightly bluer than \lbgn\ 
(with the slope $\beta$ determined directly from the observed colors) and consequently have on average lower   E(B-V) parameter, 
as derived from the spectral fitting. 
However the differences are not very large. 
An important independent check would be to assess the dust content 
e.g. from  far IR observations,
rather than through the E(B-V)  which is subject to larger uncertainties and is  strongly degenerate with the  fitted age.

\begin{table}
\caption[]{Average properties of \lbgl and \lbgn}
\begin{tabular}{llll}
\hline
\hline
Property & $LBG_L$ & $LBG_N$  & P(K-S)  \\
\hline
$N_{gal}$              &             19     & 28 &                   \\ 
z                      & 24.89$\pm0.13$     & 24.85$\pm 0.10$        & 0.99  \\
$[3.6]$                & 24.6$\pm 0.2$      & 23.8$\pm 0.15$         & 0.02 \\
$\beta$                & -2.0$\pm 0.11$     & -1.7$\pm 0.12$         & 0.32 \\
L(1400) (cgs)          & 1.1$\times 10^{29}$& 1.0$\times 10^{29}$    & 0.65 \\ 
SFR ($M_\odot yr^{-1}$)& 76$\pm34 $         &  74$\pm 22$            & 0.90 \\
Age (Myrs)             & 200$\pm 50$        & 410$\pm 70$            & 0.02 \\ 
E(B-V)                 & 0.07$\pm 0.01$     & 0.10$\pm 0.02$         & 0.58 \\
Mass ($M_\odot$) & $5\pm 1 \times 10^9$&$2.3\pm 0.8 \times 10^{10}$&$<$0.002\\
Metallicity$^{a}$      & 0.55$\pm 0.10$     & 0.48$\pm0.08$          & 1.0 \\
$\tau $                & 4.5$\pm 1.4$       & 5.5$\pm 1.3$           & 0.96 \\
Age/$\tau$             &  0.33$\pm0.13$     &  0.99$\pm0.23$         & 0.06\\
\hline 
\hline
\end{tabular}
\label{popgal}
\\
Average observational and  physical  properties for the samples 
of \lbgl\ and \lbgn\ separately: $\beta$ is the UV slope derived from the i-z color, L(1400) is the total luminosity at 1400\AA resframe; SFR is the total instantaneous star formation rate; Mass  in the total stellar mass assembled in the galaxies; $\tau$ is the star formation e-folding timescale. 
\\
 $^{a}$ We remind that this parameter is the most uncertain from the SED fit output, and that only four  values of metallicity were considered in the fitting 
procedure, see text for more details.
\end{table}

\end{document}